# THE PERMUTATION-SPECTRUM TEST: IDENTIFYING PERIODIC SIGNALS USING THE MAXIMUM FOURIER INTENSITY

B. O'NEILL,[*] *Australian National University*[**]

WRITTEN 30 NOVEMBER 2020

Abstract

This paper examines the problem of testing whether a discrete time-series vector contains a periodic signal or is merely noise. To do this we examine the stochastic behaviour of the maximum intensity of the observed time-series vector and formulate a simple hypothesis test that rejects the null hypothesis of exchangeability if the maximum intensity spike in the Fourier domain is "too big" relative to its null distribution. This comparison is undertaken by simulating the null distribution of the maximum intensity using random permutations of the time-series vector. We show that this test has a p-value that is uniformly distributed for an exchangeable time-series vector, and that the p-value increases when there is a periodic signal present in the observed vector. We compare our test to Fisher's spectrum test, which assumes normality of the underlying noise terms. We show that our test is more robust than this test, and accommodates noise vectors with fat tails.

DISCRETE FOURIER TRANSFORM; MAXIMUM INTENSITY; PERMUTATION SPECTRUM TEST.

An important problem in signal analysis is distinguishing between noise containing one or more periodic signals and noise containing no signals —i.e., when an observed times-series vector is merely noise. This is usually done by looking at the signal in Fourier space, where noise manifests asymptotically as uniform intensity over all signal frequencies, and periodic signals manifest in intensity spikes. With a finite time-series vector, there is some randomness, and leakage over signal frequencies, so we do not get perfectly uniform intensity even when it is pure noise. Nevertheless, with a sufficiently large sample size we expect that the intensity of noise in Fourier space will not contain large intensity spikes. Walker (1914) and Fisher (1929) proposed testing for the presence of a periodic signal by looking at a measure of the maximum signal intensity —i.e., the "biggest spike" in the frequency domain. Fisher implemented this spectrum test by finding the sampling distribution of a normed test statistic assuming normally distributed noise terms, drawing an analogy to the distribution of the gaps between uniformly distributed points on a circle (see Fisher 1939; 1940; Feller 1971, §III.3; Siegel 1978; 1980).

---
[*] E-mail address: ben.oneill@hotmail.com.
[**] Research School of Population Health, Australian National University, Canberra ACT 0200, Australia.

Fisher's spectrum test is a brilliant innovation in the field of signal detection, since it allows one to test for the presence of a signal without pre-specifying the signal frequency. By looking at the distribution of a normed measure of the maximum signal intensity, this "cherry-picking" of the highest evidence for a signal is accounted for in the test. Later work in local and global detection of signals has varied and extended this underlying method, leading to a range of tests for signals without pre-specification of a frequency (see e.g., Hartley 1949; Siegel 1980; Hinich 1982; Thomson 1982; Denison et al. 1999; Paparoditis 2009; Dwivedi and Rao 2011). Since classical hypothesis tests require the user to determine the null distribution of the test statistic, literature in this field generally seeks to derive an exact or asymptotic null distribution via an assumption of normally distributed noise terms.

In this paper we will argue against the approach of using parametric spectrum tests based on an assumption of normality for the noise terms. In lieu of this approach, we put forward a non-parametric permutation test analogous to Fisher's spectrum test, but without any assumption on the distribution of the error terms. Time-series models with normally distributed error terms are commonly used in practical analysis, and for many purposes —such as estimation of model coefficients— the inferences are robust to this assumption. Nevertheless, this distributional assumption is problematic in signal detection, since it rules out "fat tailed" noise distributions that give extreme noise values. We will show that even one or two extreme noise values can manifest in the Fourier domain as a large spike in intensity that makes it look like a signal is present. If one assumes that noise is normally distributed, when it actually has a fat-tailed distribution, one can dramatically underestimate the incidence of large intensity spikes.

In order to give a more robust test for the presence of a signal, we will construct a permutation test that does not require any assumption about the underlying distribution of the noise terms, save that the vector of noise terms is exchangeable. Rather than deriving the null distribution from theoretical assumptions about the underlying distribution, permutation tests use random permutation of the observed data vector to simulate the null distribution of a test statistic under the null hypothesis of exchangeability (Hoeffding 1952; Box and Anderson 1955; Edgington 1987). We consider a discrete time-series vector with $n$ observations, which we transform into Fourier space via the discrete Fourier transform (DFT). (For comprehensive details on this transform see e.g., Briggs and Henson 1995.) Following Fisher's approach, we measure the evidence for a signal using the maximum normed intensity, but we simulate the null distribution of this statistic using random permutations of the data vector.

## 1. The DFT and maximum scaled intensity of a discrete time-series vector

Let $\boldsymbol{y} = (y_1, y_2, \ldots, y_n) \in \mathbb{C}^n$ be an observable time-series vector with $n$ complex values. Our analysis will look in generality at complex time-series, but in most statistical applications that values are real numbers. For simplicity we assume that these are evenly spaced observations at times $t = 1, \ldots, n$. We denote the sample mean of the vector by $\bar{y} = \sum y_t / n$ and we take the (unitary) Discrete Fourier Transform (DFT) after removal of the sample mean, which is:

$$\mathcal{F}_{\boldsymbol{y}}(\delta) \equiv \frac{1}{\sqrt{n}} \sum_{t=1}^{n} (y_t - \bar{y}) \exp(-2i\pi\delta(t-1)) \quad \text{for all } \delta \in \mathbb{R}.$$

The DFT is a unitary periodic transformation that is often characterised by a DFT matrix and its inverse, over the fundamental frequency set $\mathfrak{F} = \{0, 1/n, \ldots, (n-1)/n\}$. These values are sufficient to characterise the DFT, so we focus on the DFT vector $\boldsymbol{r} = (r_1, r_2, \ldots, r_n)$ where $r_k = \mathcal{F}_{\boldsymbol{y}}((k-1)/n)$ denotes the DFT at the frequency $(k-1)/n$. The DFT gives a complex output for each frequency; removing the sample mean gives the fixed value $r_1 = 0$. Since each DFT is a complex value, the intensity of the signal at each frequency is given by the DFT norm $w_k \equiv \|r_k\|$, giving the intensity vector $\boldsymbol{w} = (w_1, w_2, \ldots, w_n)$.

Some basic properties of the DFT vector taken over the fundamental frequencies are shown in Appendix I. An important property is that for any initial random vector $\boldsymbol{Y} = (Y_1, Y_2, \ldots, Y_n)$ composed of uncorrelated values with a fixed mean $\mu$ and variance $\sigma^2$, the DFT values over the fundamental frequencies are uncorrelated with zero mean and variance $\sigma^2$ (except for the value at the zero frequency, which is zero). In the case where the initial time-series vector is a normal random vector, the DFT values over the fundamental frequencies are independent.

Asymptotically, each sinusoidal signal in the data manifests as a perfect intensity "spike" at its signal frequency. However, with a finite sample, even a perfect sinusoidal signal "leaks" into the surrounding signal frequencies. If we plot the DFT over the whole continuous range of frequency values (i.e., not just the fundamental frequencies), we see that a perfect sinusoidal signal manifests in a main hump at its frequency value, with diminishing oscillations rippling out from that frequency. (The size of these oscillations diminishes as the sample size increases, and vanishes in the limit.) Once we restrict attention to the fundamental frequencies in the DFT vector, the oscillations manifest as a smoothly diminishing intensity spike. In any case, the phenomena of bleeding of signals into other frequencies is known as "leakage".

The intensity is affected by the variance of the observed time-series vector. In order to get a measure of signal intensity that is comparable across different time-series vectors, we norm our intensity measure to remove the effect of the variance. If we denoting the sample variance by $s^2 = \sum |y_t - \bar{y}|^2/(n-1)$ then Parseval's theorem tells us that:

$$\sum_{k=1}^{n} w_k^2 = \sum_{k=1}^{n} \|r_k\|^2 = \sum_{t=1}^{n} |y_t - \bar{y}|^2 = (n-1)s^2.$$

Since the norm of the intensity vector is related to the variance, we will facilitate our analysis be defining standardised and studentised versions of the intensity, given respectively by:

$$\text{SI}_k(\boldsymbol{y}) \equiv \frac{w_k}{\sigma} = \frac{\|r_k\|}{\sigma} \qquad \widehat{\text{SI}}_k(\boldsymbol{y}) \equiv \frac{w_k}{s} = \frac{\|r_k\|}{s}.$$

The value $\widehat{\text{SI}}_k(\boldsymbol{y})$ gives scaled measure of intensity of the signal at the frequency $(k-1)/n$, since it filters out the effect of the sample variance. For an initial time-series vector $\boldsymbol{y}$ we can look at the vector $\widehat{\textbf{SI}}(\boldsymbol{y}) = (\widehat{\text{SI}}_k(\boldsymbol{y})|k = 1, ..., n)$ with the scaling property $\|\widehat{\textbf{SI}}(\boldsymbol{y})\|^2 = n-1$ (i.e., the squared-norm of the vector is equal to the number of degrees-of-freedom after subtraction of the mean from the time-series vector). This scaled intensity measure filters out the effects of the length and variability of the initial time-series vector, leaving us with a measure that reflects only the intensity of the periodic signal at the input frequency.

Since intensity spikes in the frequency domain constitute evidence of the presence of a signal, it is natural to look at the largest spike to assess whether it is "too big" to plausibly be caused by random variation within a set of exchangeable values. We focus on the maximum scaled intensity spike over the fundamental frequencies, which is $\widehat{\text{MSI}}(\boldsymbol{y}) \equiv \max_{k \in \mathbb{Z}} \widehat{\text{SI}}_k(\boldsymbol{y})$. This is the approach taken in the original spectrum test by Fisher (1929).[1] Since we remove the sample mean of the time-series vector prior to taking the DFT, the zero frequency has zero intensity, so we can exclude this point from consideration.[2] By taking the maximum intensity only over the fundamental frequencies we take advantage of the fact that the DFT values are uncorrelated over these frequencies in a wide class of cases.

---

[1] Our test statistic is slightly different to the one used in Fisher (1929), but they are monotonically related. In our notation, Fisher used the statistic $\widehat{\text{MSI}}(\boldsymbol{y})^2/(n-1)$ which is a scaled version of the maximum squared-intensity (so that the vector sums to unity). Since these are monotonically related test statistics, our usage does not change the test; our usage is chosen because we find our test statistic to be more easily interpreted and more easily comparable across different time-series vectors than the one used by Fisher.

[2] When $\boldsymbol{y}$ is a real vector (i.e., not complex) the DFT values are reflected via complex conjugation around the Nyquist frequency $\delta = \frac{1}{2}$, so the intensities are also reflected around the Nyquist frequency. In this case the initial time-series vector is fully characterised by the DFT over the frequencies $\mathfrak{G} = \{\delta \in \mathfrak{F} | 0 < \delta \leq \frac{1}{2}\}$.[2] Thus, if we confine our attention only to real time-series vectors, we can ignore half of the DFT output.

The scaled intensity values are used primarily to detect signals with fixed frequency, but they also relate to the sample auto-correlation in the time-series vector, which means that they also relate to quasi-periodic signals induced by auto-correlation in the series (i.e., signals that have a varying frequency). To demonstrate this, we will use the sample auto-covariances:

$$\hat{\gamma}_\ell \equiv \frac{1}{n-\ell-1}\sum_{t=1}^{n-\ell}(y_t - \bar{y})(y_{t+\ell} - \bar{y}) \quad \text{for all } \ell = 0,1,2,\ldots,n-2.$$

Using these values, the squared intensity values can be written as:

$$\|\mathcal{F}_y(\delta)\|^2 = \frac{1}{n}\left[\left(\sum_{t=1}^{n}(y_t - \bar{y})\cos(-2\pi\delta(t-1))\right)^2 + \left(\sum_{t=1}^{n}(y_t - \bar{y})\sin(-2\pi\delta(t-1))\right)^2\right]$$

$$= \frac{1}{n}\left[\sum_{t=1}^{n}\sum_{s=1}^{n}(y_t - \bar{y})(y_s - \bar{y})\binom{\cos(-2\pi\delta(t-1))\cos(-2\pi\delta(s-1))}{+\sin(-2\pi\delta(t-1))\sin(-2\pi\delta(s-1))}\right]$$

$$= \frac{1}{n}\left[\sum_{t=1}^{n}\sum_{s=1}^{n}(y_t - \bar{y})(y_s - \bar{y})\cos(2\pi\delta(t-s))\right]$$

$$= \frac{1}{n}\left[\sum_{t=1}^{n}(y_t - \bar{y})^2 + 2\sum_{\ell=1}^{n-1}\cos(2\pi\ell\delta)\sum_{t=1}^{n-\ell}(y_t - \bar{y})(y_{t+\ell} - \bar{y})\right]$$

$$= \frac{n-1}{n}\cdot\hat{\gamma}_0 + 2\sum_{\ell=1}^{n-2}\frac{n-\ell-1}{n}\cdot\hat{\gamma}_\ell\cdot\cos(2\pi k\delta).$$

Using the autocorrelation function $\hat{\rho}_\ell = \hat{\gamma}_\ell/\hat{\gamma}_0$, and noting that $\hat{\gamma}_0 = s^2$ is the sample variance, we then have the squared scaled intensity:

$$\frac{\|\mathcal{F}_y(\delta)\|^2}{s^2} = \frac{n-1}{n} + 2\sum_{\ell=1}^{n-2}\frac{n-\ell-1}{n}\cdot\hat{\rho}_\ell\cdot\cos(2\pi\ell\delta).$$

Letting $\phi \equiv \cos(2\pi\delta)$ we have $\cos(2\pi\ell\delta) = T_\ell(\cos(2\pi\delta)) = T_\ell(\phi)$ where the function $T_\ell$ is the $\ell$th Chebyshev polynomial of the first kind (see e.g., Mason and Handscomb 2002). This allows us to write the squared scaled intensity in terms of $\phi$ as:

$$\frac{\|\mathcal{F}_y(\delta)\|^2}{s^2} = \frac{n-1}{n} + 2\sum_{\ell=1}^{n-2}\frac{n-\ell-1}{n}\cdot\hat{\rho}_\ell\cdot T_\ell(\phi).$$

As can be seen, the scaled intensity is a sum of terms which are multiples of polynomial terms in $\phi$ multiplied by the sample auto-correlation values. If the true auto-correlation values are zero at all lags then the sample auto-correlations will converge to zero and so the individual terms in the sum become small as $n \to \infty$. This is a general form that holds at any frequency, so it also applies on the fundamental frequencies.

## 2. Testing for a periodic signal — the permutation-spectrum test

We now develop a hypothesis test to test for the presence of one or more periodic signals in a time-series vector. The null hypothesis is that the time-series vector $\boldsymbol{Y}$ is exchangeable. (Note that a sufficient condition for exchangeability is for the values in the time series vector to be IID from some underlying marginal distribution.) The alternative hypothesis is that this time-series vector contains one or more non-constant periodic signals, so that its elements are not exchangeable. The test statistic $\widehat{\text{MSI}}(\boldsymbol{y})$ gives the strongest intensity in the frequency domain, so it gives a measure of evidence for a periodic signal — a larger value for this statistic constitutes greater evidence in favour of the alternative hypothesis.

Let $F_0$ denote the true null distribution of the maximum scaled intensity, which depends on the distribution of the underlying time-series vector. We will not assume any particular form for this underlying distribution, and so we refrain from attempting a theoretical derivation of the null distribution based on a stipulated distribution for the time-series vector. Instead, our test estimates the true null distribution by simulation. To do this, let $\boldsymbol{\Pi} = (\Pi_1, \ldots, \Pi_n)$ be a random permutation that reorders the time indices $(1, \ldots, n)$, with every possible order being equally probable. Let $\boldsymbol{y}_{\boldsymbol{\Pi}} = (\boldsymbol{y}_{\Pi_1}, \ldots, \boldsymbol{y}_{\Pi_m})$ denote the random vector that randomly permutes the time-series $\boldsymbol{y}$. We simulate $M$ random permutations $\boldsymbol{\Pi}_1, \ldots, \boldsymbol{\Pi}_M$ and compute the maximum scaled intensity $\widehat{\text{MSI}}(\boldsymbol{y}_{\boldsymbol{\Pi}_m})$ for each permutation of the time-series vector. We can then estimate the null distribution using the empirical distribution of these values:

$$\hat{F}(s) = \frac{1}{M} \sum_{m=1}^{M} \mathbb{I}(\widehat{\text{MSI}}(\boldsymbol{y}_{\boldsymbol{\Pi}_m}) \leq s) \qquad \text{for all } s \in \mathbb{R}.$$

We know that a permutation of an exchangeable random vector does not change its distribution. Since $\widehat{\text{MSI}}(\boldsymbol{Y}) \sim F_0$ it follows that —under the null hypothesis of exchangeability— we also have $\widehat{\text{MSI}}(\boldsymbol{Y}_{\boldsymbol{\Pi}_m}) \sim F_0$ which gives $\mathbb{I}(\widehat{\text{MSI}}(\boldsymbol{y}_{\boldsymbol{\Pi}_m}) \leq s) \sim \text{Bern}(F_0(s))$. This gives us:

$$\hat{F}(s) \sim \frac{1}{M} \cdot \text{Bin}(M, F_0(s)),$$

Since $\hat{F} \to F_0$ as $M \to \infty$ (almost sure uniform convergence, via the Glivenko-Cantelli lemma), the simulated null distribution is a strongly consistently estimator of the true null distribution. In particular, we can estimate the true p-value of the test as:

$$\hat{p}(\boldsymbol{y}, \boldsymbol{\Pi}_1, \ldots, \boldsymbol{\Pi}_M) = \frac{1}{M} \sum_{m=1}^{M} \mathbb{I}(\widehat{\text{MSI}}(\boldsymbol{y}_{\boldsymbol{\Pi}_m}) \geq \widehat{\text{MSI}}(\boldsymbol{y})).$$

The simulated p-value is an unbiased estimator of the true p-value of the test. If necessary, we can appeal to the Wilson score interval (Wilson 1927) to obtain an appropriate confidence interval for the true p-value. By setting $M$ to be a large number of simulations we can obtain an accurate estimate of the true null distribution and the corresponding p-value for the test (computational issues are discussed in Pagano and Trictchler 1983). Each simulation applies a permutation to the time-series vector, computes its DFT and resulting scaled intensity vector, and computes the maximum scaled intensity under the simulation.

In Figure 1 below we show a Permutation Spectrum Test Plot illustrating the test. This plot is obtained from a noise vector composed of $n = 60$ IID values from a T-distribution with two degrees-of-freedom. The bar-plot on the left shows the scaled intensity of the time-series over the frequencies in the Nyquist range, and the violin plot on the right shows the simulated null density for the maximum scaled intensity, with quartiles shown as horizontal lines. The plots share their vertical axis, and both show the maximum scaled intensity of the observed time-series vector is shown as a red dot. By looking at the red dot falls within the simulated density we can see whether there is evidence to reject the null hypothesis in favour of the alternative. In the particular case shown in the Figure, the simulated p-value is $\hat{p}(\boldsymbol{y}) = 0.7175$, so there is no evidence to reject the null hypothesis.

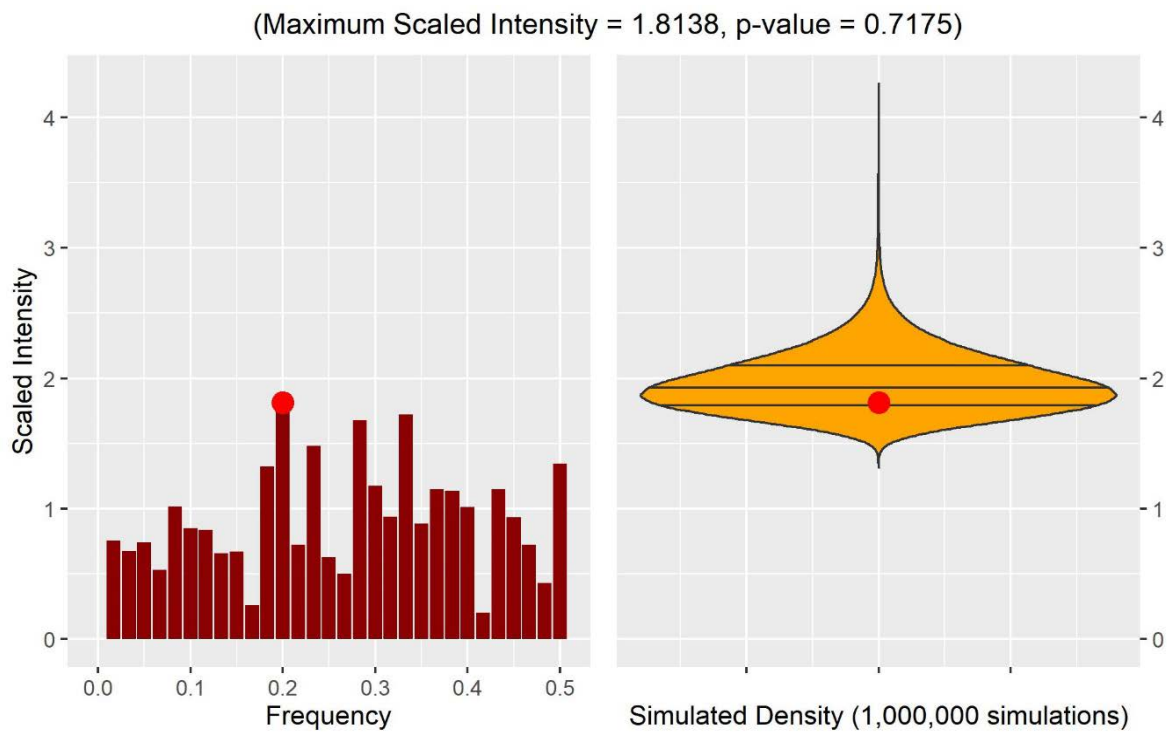

**Figure 1:** Permutation-Spectrum Test Plot (noise vector)

In a later section we conduct a simulation study looking at the power of the test in detecting periodic signals. For present purposes we illustrate this with a single example. In Figure 2 we show the Permutation Spectrum Test Plot for a time-series vector taking the same time-series vector as in the previous plot, but with an added sinusoidal signal constructed so that the signal-to-noise ratio is unity (i.e., the signal is the same "magnitude" as the noise).[3] The simulated p-value is $\hat{p}(\mathbf{y}) = 0.0018$, so there is strong evidence of a signal. In this case the test correctly detects a periodic signal in the noise.

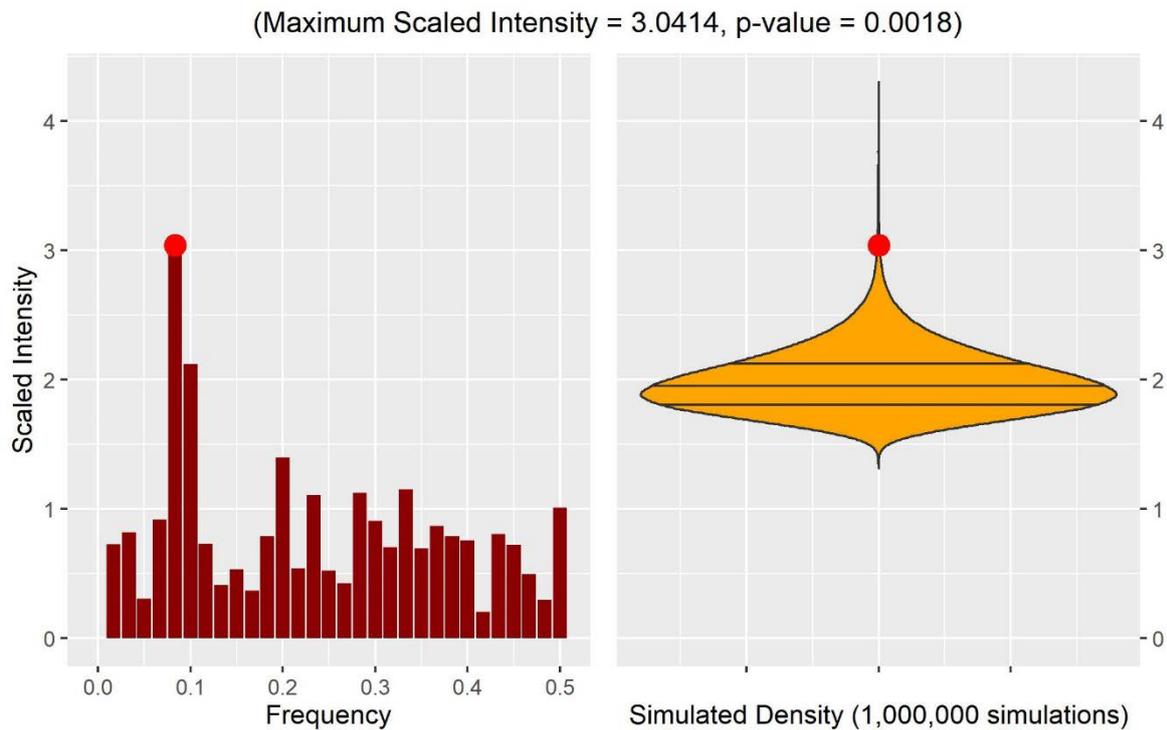

**Figure 2:** Permutation-Spectrum Test Plot (signal + noise vector)

By construction, the p-value for our test is uniformly distributed under the null hypothesis.[4] If $\mathbf{y}$ is exchangeable then the values $\text{MSI}(\mathbf{y}), \text{MSI}_1, \ldots, \text{MSI}_M$ must occur in a random order. If $\mathbf{y}$ is absolutely continuous then there can be no ties in these values, so we obtain:

$$\hat{p}(\mathbf{y}) \sim \text{U}\left\{0, \frac{1}{M}, \frac{2}{M}, \ldots, \frac{M-1}{M}, 1\right\}.$$

---

[3] For this ratio, the "magnitude" of the signal vector and noise vector were measured by the sum of the absolute value of their elements.

[4] As is standard in hypothesis tests, if $\mathbf{y}$ is discrete then the p-value will not be exactly uniformly distributed. In this case there can be ties in the simulated values of the maximum scaled intensity, and ties between these values and the observed maximum scaled intensity.

The distribution of the p-value under the alternative hypothesis is much more complicated, but it will tend to be stochastically reduced when there is a periodic signal in the time-series vector. The intuition for why this occurs is simple. We know that exchangeable noise manifests in the frequency domain in a manner that is stochastically uniform, not favouring any frequencies. Contrarily, a periodic signal manifests in the frequency domain as an uneven set of spikes that are concentrated on its most intense frequencies. In the scaled intensity, the addition of these uneven intensity spikes from the signal will increase some of the scaled intensities and decrease others. The resulting systematic unevenness of the intensity spikes from the signal will tend to increase the maximum scaled intensity, since they are concentrated over particular frequencies. However, once we impose a random permutation to mimic the null hypothesis, the periodic signal that was in the time-series becomes "muddled" and no longer favours any frequencies. Thus, for a time-series vector containing a periodic signal, the observed maximum scaled intensity will tend to be larger than under our simulation of the null hypothesis.

**3. Simulating the permutation spectrum test — power analysis**

The above analysis establishes that the p-value of the permutation spectrum test is uniformly distributed under the null hypothesis of an exchangeable time-series vector. We now perform a simulation analysis looking at the distribution of the p-value when there is a periodic signal in the time-series vector. There are three variable inputs that are of interest in this simulation analysis: (1) the shape of the underlying noise distribution; (2) the signal-to-noise ratio in the time series; and (3) the number of observations in the time series vector. We will generate a large number of tests varying these three inputs so that we can illustrate the power of the test to detect signals of various magnitudes in noise sequences of various kinds. To simplify our analysis we will compare only two noise distributions; a normal distribution with "thin tails" and a T distribution with two degrees-of-freedom with "fat tails".

We begin by giving our analysis some additional structure, to allow us to reframe our null and alternative hypotheses in quantitative terms. In all our simulations in this section we generate a random sinusoidal signal $\boldsymbol{s}$ and a random noise vector $\boldsymbol{\varepsilon}$ each with the same "magnitude" (the sum of absolute values of their elements). We take a value $\lambda \neq 0$ for the signal-to-noise ratio and generate the observed time-series vector:

$$\boldsymbol{y} \equiv \lambda \cdot \boldsymbol{s} + \boldsymbol{\varepsilon},$$

The noise vector is generated by taking IID elements from some chosen noise distribution, and the signal vector is a perfect sinusoidal signal with a random frequency in the Nyquist range.[5] *Ceteris paribus*, the larger the value of $\lambda$ the more prominently the signal will stand out from the noise, and the larger the value of $n$ the greater our capacity to detect a signal. Within the confines of this model form, the noise vector $\boldsymbol{\varepsilon}$ is exchangeable and the signal vector $\boldsymbol{s}$ is non-exchangeable, so the hypotheses reduce to the quantitative form:

$$H_0: \lambda = 0 \qquad H_A: \lambda \neq 0.$$

In Figure 3 below we show the p-values from simulated permutation spectrum tests performed on random time-series vectors using noise from a standard normal distribution. In Figure 4 we perform the same analysis using noise from a Student's T distribution with two degrees-of-freedom. Under each analysis, for each time-series length $n$ and each signal-to-noise ratio we generated $K = 10000$ random noise vectors and signal vectors[6] and performed the test on the combined time-series vector composed of the signal and noise. (In the case where the signal-to-noise ratio was zero there was no signal.) The figures show jitter-plots of simulated p-values and corresponding boxplots of those p-values for each combination of the time-series length $n$ and the signal-to-noise ratio $\lambda$.

Both figures show the same general pattern in the distribution of the p-values as we alter the time-series length and signal-to-noise ratio. With no signal the p-value is distributed uniformly (something we have already shown deductively), but as we increase the signal-to-noise ratio the p-value decreases, showing evidence of a signal. As expected, tests on longer time-series have greater power, shown by the fact that the simulated p-values decrease faster with respect to the signal-to-noise ratio when $n$ is larger. As is to be expected, the rate at which this occurs differs between Figures 3-4, based on the shape of the underlying distribution of the noise. If we use a noise distribution with "fatter tails" then there will be noise values of high magnitude, and these can falsely appear to be a signal if they happen to occur roughly in periodic intervals. This means that it is harder to detect periodic signals in noise with fatter tails, and so we see that the p-value decreases more rapidly in Figure 3 than in Figure 4.

---

[5] The randomness in the sinusoidal signal comes from generating a random frequency over the Nyquist range. The phase-angle does not affect the intensity so it is set to zero. The amplitude is set by the requirement of equal magnitude with the noise vector.

[6] For each of the two noise distributions in Figures 3-4 we used four different vectors lengths and six different values of the signal-to-noise-ratio, so each figure contains $4 \times 6 \times 10{,}000 = 240{,}000$ simulated p-values.

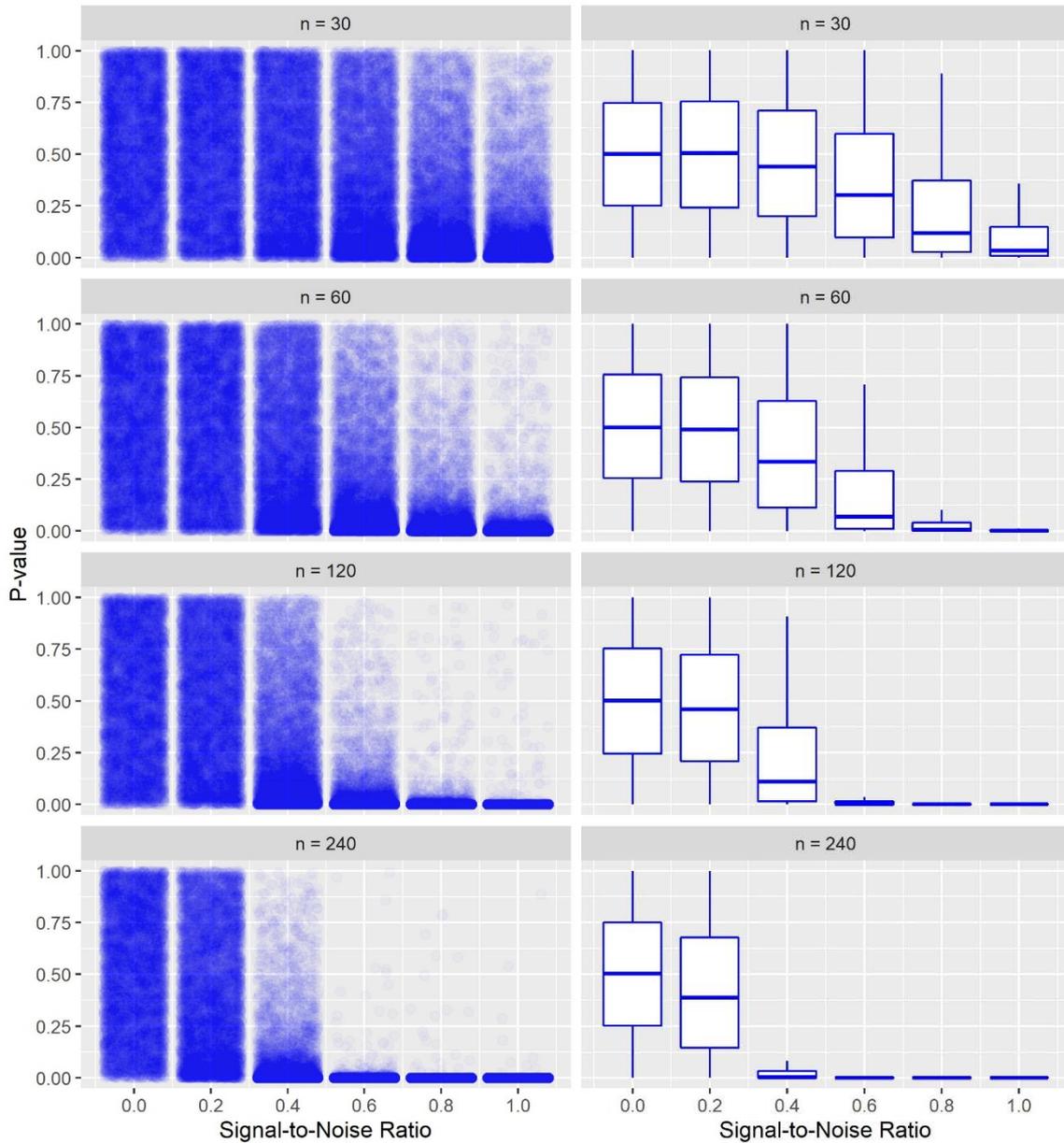

**Figure 3:** Simulated Permutation-Spectrum Tests
(noise vector simulated from standard normal distribution)

For the analysis shown in Figure 3 we used noise from a standard normal distribution and so the noise values did not have values with particularly high magnitude. Based on the boxplots of the p-values, we can see that with $n = 60$ time-series values the test reliably detects signals with a signal-to-noise ratio of $\lambda = 0.8$, with $n = 120$ time-series values the test reliably detects signals with a signal-to-noise ratio of $\lambda = 0.6$, and with $n = 240$ time-series values the test reliably detects signals with a signal-to-noise ratio of $\lambda = 0.4$. The power of the test at a particular significance level can be estimated from the simulations, but we focus here on the distribution of the p-values rather than comparing them to a particular significance level.

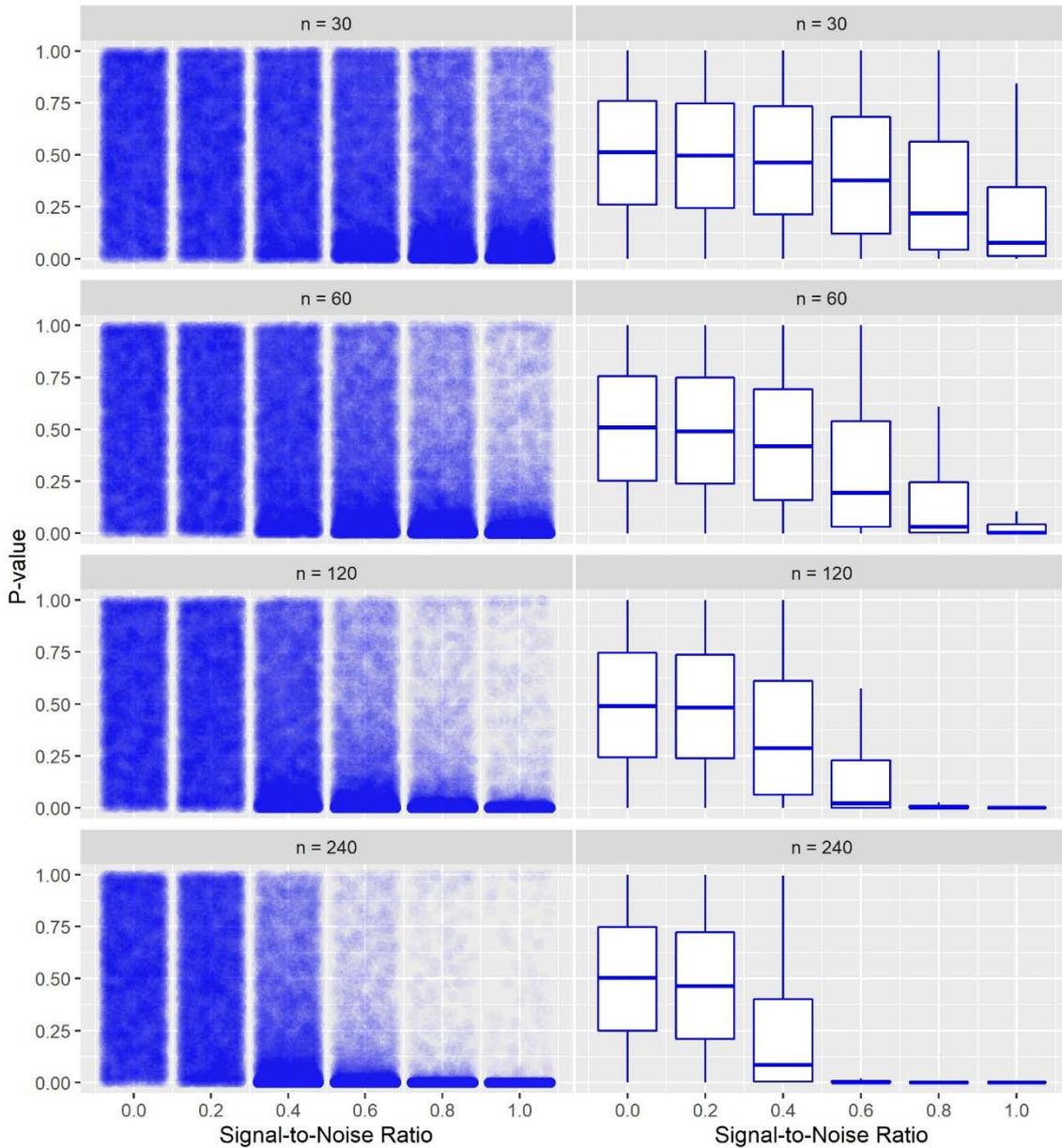

**Figure 4:** Simulated Permutation-Spectrum Tests
(noise vector simulated from T distribution with two degrees-of-freedom)

For the analysis shown in Figure 4 we used noise from a T distribution with two degrees-of-freedom and so there were some higher magnitude noise values. Based on the boxplots of the p-values, we can see that with $n = 60$ time-series values the test reliably detects signals with a signal-to-noise ratio of $\lambda = 1.0$, with $n = 120$ time-series values the test reliably detects signals with a signal-to-noise ratio of $\lambda = 0.8$, and with $n = 240$ time-series values the test reliably detects signals with a signal-to-noise ratio of $\lambda = 0.6$. In this case there is lower power than when we used noise from a standard normal distribution. This occurs because large noise values can look like signals when they appear in periodic patterns.

For any given value of the signal-to-noise ratio $\lambda$ and the significance level $\alpha$, the true power of the test is $\text{Power}_\alpha(\lambda) \equiv \mathbb{P}(\text{Reject } H_0|\lambda, \alpha) = \mathbb{P}(p(Y) \leq \alpha|\lambda)$. We can estimate the power over the inputs used in our analysis by computing the proportion of our simulated p-values that fall below the stipulated significance level. In Figure 5 below we show the estimate power for tests conducted at the $\alpha = 0.05$ significance level. (The estimated values used in this plot are shown in Table 1 in the Appendix.) The true power at $\lambda = 0$ is the significance level $\alpha = 0.05$ but our simulated estimates vary slightly from this. As we increase the signal-to-noise ratio $\lambda$ or the time-series length $n$ we see that the power increases. This occurs more rapidly when the noise was generated from a normal distribution (thinner tails) than when it was generated from a T distribution with two degrees-of-freedom (fatter tails).

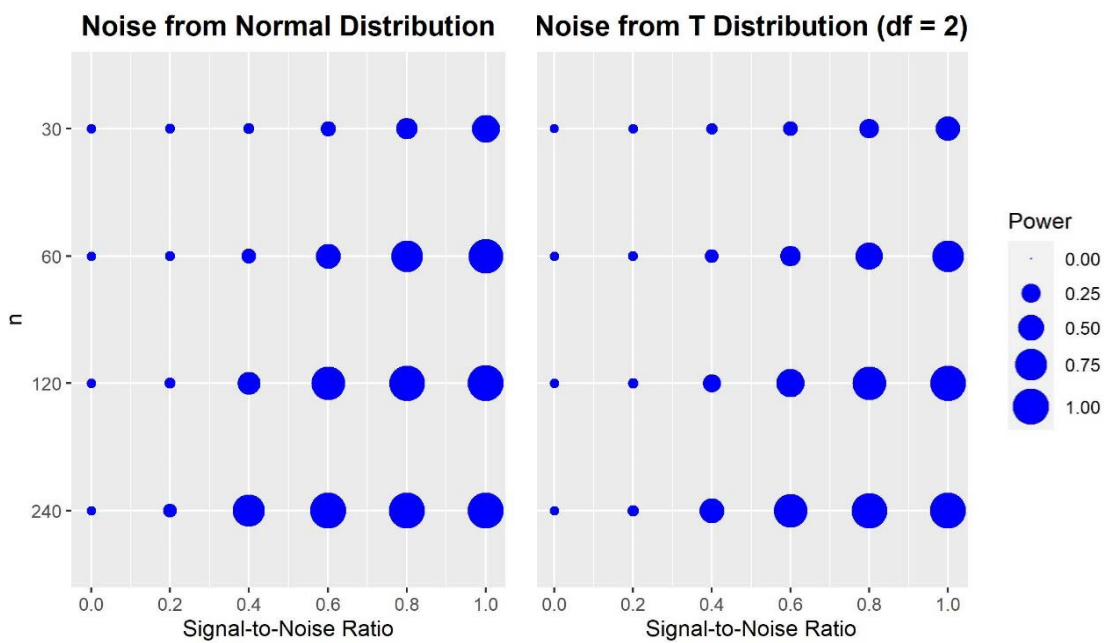

**Figure 5:** Estimated Power of Permutation-Spectrum Tests at $\alpha = 0.05$
(noise vector simulated from T distribution with two degrees-of-freedom)

Our simulation analysis here has examined the distribution of the p-value —and the resulting power of the test at a given significance level— taken over two different noise distributions and a range of values of the observed time-series length and the signal-to-noise ratio. It is worth noting that the periodic signal in our analysis is a perfect sinusoidal signal, which is the easiest type of signal to detect (since it concentrates on a single frequency with no harmonics). Intuitively we would expect other periodic signals that are not perfectly sinusoidal to be harder to detect, and therefore these would tend to have p-values that decrease less rapidly in $n$ and $\lambda$.

## 4. Discussion and conclusion

The permutation-spectrum test formulated here provides a non-parametric test for the presence of a signal in an exchangeable noise vector. The test and accompanying plot are implemented in `R` in the `ts.extend` package (O'Neill 2020) by the `spectrum.test` function and the `plot` function for the resulting test object. The function takes an input signal and a specified number of simulations for the test and produces a test object that can be printed or plotted. (All the computations of the permutation-spectrum test in this paper were done using this function; Figures 1-2 were produced directly by printing the test object from this function.)

The permutation-spectrum test has several potential applications in time-series analysis, either as a limited diagnostic tool for analysing residual vectors, or as a broader tool to determine the presence of signals for a fitted model. In the latter case, the analyst can begin with an initial time-series vector and conduct the test to see if there is a signal present; if so, the analyst fits a sinusoidal signal to the data and removes the estimated signal to give a residual vector. The residual vector is then tested, and the process repeats until there are no remaining signals. By this method, the analyst estimates the number of signals for the purposes of model construction. Alternatively, an analyst can use another method to determine an appropriate model (e.g., estimating the number of signal terms) and use the test only to confirm that the residual vector does not have any remaining signals.

An alternative method to the use of the permutation spectrum test is to construct a nested model for signals in noise, with a general model form that accommodates any number of signals. The analyst can fit the model under any stipulated number of signals and then use goodness-of-fit metrics or likelihood-ratio tests to determine the appropriate number of signals in the model. These existing methods are all parametric in nature, and so they hinge on specification of the model form. In particular, such model will almost always stipulate a family of distributions for the noise terms, and this means that the methods are subject to poo performance if there is misspecification of the noise distribution, and the true noise distribution is more (or less) likely to give high-magnitude terms that can be mistaken for signals when they occur in a periodic pattern.

One major advantage of the permutation spectrum test is that the null hypothesis is broad enough to accommodate IID values from any underlying noise distribution. Any noise vector composed of IID values is exchangeable, and it falls within the purview of our null hypothesis. This means that the test automatically accommodates cases where the noise has high kurtosis. We do not test for any particular noise distribution, and in particular, we do not test only for normally distributed "white noise". If normality of the time-series vector is at issue, the analyst can of course employ other tests for this purpose, but our test is designed only to distinguish between exchangeable observations, versus observations that contain a periodic signal. As such, our test is robust to the underlying distribution of a noise vector.

It is important to stress that the permutation-spectrum test does not provide a substitute for modelling time-series with periodic signals. Its main function is to augment these models by providing a simple method to test a residual vector to see if there is evidence of any remaining signal. We recommend the test as a useful diagnostic tool for time-series analysis, which can be routinely applied to the residual vector from a time-series model. It can also be used more broadly to assist in model formulation, through estimation of the number of sinusoidal signals in data.

## Supplementary Materials

The code for our simulation analysis and the resulting arrays of simulated p-values in this paper are available in the supplementary materials:

```
PST Simulation.Rmd            R Markdown file for simulation analysis
TESTS.normdist (10000).rds    Array of simulated p-values (10,000 simulations)
                              (noise distributed by normal distribution)
TESTS.Tdist (10000).rds       Array of simulated p- values (10,000 simulations)
                              (noise distributed by T distribution with df = 2)
```

# Appendix I: Properties of the DFT vector

Here we give a brief derivation of some properties of the DFT vector of an observable time-series vector $\boldsymbol{y} = (y_1, \ldots, y_n)$. To facilitate our analysis we let $\omega_n \equiv \exp(-2i\pi/n)$ denote the primitive $n$th root of unity and we use the unitary DFT matrix $\mathbf{U}$ with elements:

$$U_{\ell,k} \equiv \frac{1}{\sqrt{n}} \cdot \omega_n^{(\ell-1)(k-1)}.$$

The unitary DFT matrix is an orthonormal matrix with inverse $\mathbf{U}^*$ —i.e., its conjugate transpose is its inverse (see e.g., Briggs and Henson 1995). If we let $\mathbf{C}$ denote the centering matrix then the DFT vector over the fundamental frequencies is the vector $\boldsymbol{r} = \mathbf{U}\mathbf{C}\boldsymbol{y}$. Note that we have $r_1 = 0$ in the DFT vector by virtue of the centering of the time-series vector prior to transformation into Fourier space.

The centering matrix is a symmetric idempotent matrix that can be written in spectral form as $\mathbf{C} = \mathbf{U}^*\boldsymbol{\Lambda}\mathbf{U}$, where $\boldsymbol{\Lambda}$ is a diagonal eigenvalue matrix with eigenvalues $\lambda_\ell = \mathbb{I}(\ell \neq 0)$. For any real data vector $\boldsymbol{Y}$ composed of uncorrelated values with a fixed finite mean and variance we have the moments $\mathbb{E}(\boldsymbol{Y}) = \mu\mathbf{1}$ and $\mathbb{V}(\boldsymbol{Y}) = \sigma^2\boldsymbol{I}$. Thus, with a bit of matrix algebra, we have:

$$\mathbb{E}(\boldsymbol{R}) = \mathbb{E}(\mathbf{U}\mathbf{C}\boldsymbol{Y}) = \mathbf{U}\mathbf{C}\mathbb{E}(\boldsymbol{Y}) = \mu\mathbf{U}\mathbf{C}\mathbf{1} = \mu\mathbf{U}\mathbf{0} = \mathbf{0},$$

$$\mathbb{V}(\boldsymbol{R}) = \mathbb{V}(\mathbf{U}\mathbf{C}\boldsymbol{Y}) = \mathbf{U}\mathbf{C}\mathbb{V}(\boldsymbol{Y})\mathbf{C}\mathbf{U}^* = \sigma^2\mathbf{U}\mathbf{C}\mathbf{C}\mathbf{U}^* = \sigma^2\mathbf{U}\mathbf{C}\mathbf{U}^* = \sigma^2\boldsymbol{\Lambda}.$$

As previously noted, the first element of the DFT vector is zero, since it is the DFT at the zero frequency (and is therefore set to zero by the centering of the values). The remaining values in the DFT vector are uncorrelated complex random variables with zero mean and variance $\sigma^2$. This establishes that the values in the DFT vector are uncorrelated for any underlying vector composed of uncorrelated values with fixed mean and variance.

# Appendix II: Table of Power Simulations

|  | Noise: Standard normal distribution | | | | |
|---|---|---|---|---|---|
|  | $\lambda = 0.2$ | $\lambda = 0.4$ | $\lambda = 0.6$ | $\lambda = 0.8$ | $\lambda = 1$ |
| $n = 30$ | 0.0541 | 0.0739 | 0.1576 | 0.3403 | 0.5684 |
| $n = 60$ | 0.0579 | 0.1486 | 0.4496 | 0.7682 | 0.9201 |
| $n = 120$ | 0.0743 | 0.3863 | 0.8506 | 0.9793 | 0.9925 |
| $n = 240$ | 0.1228 | 0.7868 | 0.9935 | 0.9961 | 0.9974 |
|  | Noise: Student's T distribution with two degrees-of-freedom | | | | |
|  | $\lambda = 0.2$ | $\lambda = 0.4$ | $\lambda = 0.6$ | $\lambda = 0.8$ | $\lambda = 1$ |
| $n = 30$ | 0.0533 | 0.0778 | 0.1432 | 0.2663 | 0.4354 |
| $n = 60$ | 0.0596 | 0.1188 | 0.2990 | 0.5485 | 0.7624 |
| $n = 120$ | 0.0604 | 0.2253 | 0.5826 | 0.8436 | 0.9480 |
| $n = 240$ | 0.0792 | 0.4418 | 0.8516 | 0.9653 | 0.9871 |

Note: We have omitted the simulated values of the power for $\lambda = 0$. The true power in this case is $\alpha = 0.05$. Our estimates differ slightly from the true value due to random variation in our simulations.

Table 1: Estimated Power of Permutation Spectrum Tests
(for a test at significance level $\alpha = 0.05$)